\documentclass{cargese}\usepackage{graphicx}

\let\footnote\savefootnote
\let\footnotetext\savefootnotetext 
\let\footnoterule\savefootnoterule 
\normallatexbib

\begin{document}

\articletitle[Cubic Free Field Theory]
{Cubic Free Field Theory}

\chaptitlerunninghead{Cubic Free Field Theory}

\author{Boris Pioline}

\affil{LPTHE, Universit\'es Paris 6 et 7, \\
4 place Jussieu, 75252 Paris cedex, France}    

\email{pioline@lpthe.jussieu.fr}


\begin{abstract}
We point out the existence of a class of non-Gaussian yet 
free ``quantum field theories'' in 0+0 dimensions, based on a cubic
action classified by simple Lie groups. A ``three-pronged'' version
of the Wick theorem applies. \hfill (LPTHE-P03-01, {\tt hep-th/0302043})
\end{abstract}



With the exception of a few integrable models in low dimension, much
of what we know of quantum field theory 
relies on perturbation around a free theory
with a Gaussian action $S=\phi^t Q \phi/2$. 
Such a theory is free in the sense that the generating
functional of connected diagrams
\begin{equation}
\label{1pi}
e^{\frac{i W(J)}{\hbar}}=\int [d\phi_i] \exp\left[ \frac{i}{\hbar}
 \left( S(\phi) 
+ \phi^t J\right) \right]
\end{equation}
is given {\it exactly} by the semiclassical, or saddle point, approximation 
to the integral above,
\begin{equation}
\label{qclas}
e^{\frac{i W(J)}{\hbar}}
= \frac{1}{\sqrt{\det(Q)}}  \exp\left( -\frac{i}{2\hbar} J^t Q^{-1} J \right)
\end{equation}
The quantum effective action, defined as the Legendre transform of
$W(J)$, is therefore identical to the classical
action. Interactions can then be introduced by perturbing 
this Gaussian point by higher powers of the field $\phi$; their
effects are computed efficiently by use of Wick's theorem, which 
follows immediatly from Eq.~(\ref{qclas}).

One may think that the exactness of the semiclassical approximation be
a defining property of Gaussian theories. Our goal in this gong-note
is to point out that there exist non-Gaussian theories which satisfy
the same property. Indeed, consider the simple ``action'' of four
variables, $S=C~\phi_1 \phi_2 \phi_3/ \phi_0$. As first observed 
by Kazhdan in \cite{ka}, the Legendre transform 
of $S(\phi)$ in all variables reproduces 
a similar function again,
\begin{equation}
\label{leg}
\langle\  C\frac{\phi_1 \phi_2 \phi_3}{\phi_0} 
+ \phi^t J \ \rangle_{\{\phi_i\}} = \frac{J_1 J_2 J_3}{C~J_0}
\end{equation}
The ``path'' integral (\ref{1pi}) has therefore a single saddle
point whose value is the same as the classical action up to an
inversion of $C$. Furthermore, an elementary computation
using  the Fourier transform of Dirac's delta function
repetitively shows that, for an appropriate choice of the integration measure, 
the exact generating functional $W$ is equal to its saddle point approximation,
\begin{equation}
\label{eleg}
\int \frac{[d\phi_{0123}]}{\phi_0} \exp\left[
\frac{i}{\hbar} \left( 
C \frac{\phi_1 \phi_2 \phi_3}{\hbar \phi_0} + \phi^t J \right)\right]
=
 \frac{\hbar^2}{J_0 C}
 \exp\left[ \frac{i}{\hbar}  \frac{J_1 J_2 J_3}{C~J_0} \right]
\end{equation}
After a further Legendre transform, the quantum effective action 
is therefore equal to the classical action, as in the case of
a Gaussian action.

In view of this small miracle, one may ask whether other non-Gaussian
actions give rise to free quantum theories as well\footnote{Earlier
discussions of self-reciprocal functions have been brought to
my attention, see e.g \cite{Scharnhorst} and references [75-80]
therein.}.
A necessary
requirement is that there should exist exactly one saddle point 
for any current $J$, hence the differential $dS$ should be of index
one. A possible choice is to take $S(\phi_i)$ to be 
an homogeneous polynomial of 
degree 3, divided by an extra variable $\phi_0$ so that the total
degree is 2. One is thus led to the classification of 
cubic forms $I_3(\phi_{i=1\dots n})=C_{ijk}\phi_i \phi_j \phi_k/6$ 
such that the Fourier transform of $\exp(i I_3(\phi)/\phi_0)$ 
be equal to  $\exp(i I_3(J)/J_0)$, possibly up to a one-loop
measure factor. In particular, the Legendre transform of
$I_3(\phi)/\phi_0$ should be equal to  $I_3(J)/J_0$,
as in our example (\ref{leg}) above. 

This problem has been solved recently in a more general context by Etingof, 
Kazhdan and Polishchuk \cite{ekp}. They found that the cubic forms
$I_3(\phi)$ satisfying the property above fall into one of six
cases, labeled by simple Lie groups,
\begin{enumerate}
\item[(i)] $D_{n\geq 4}: I_3 = \phi_1 (\phi_2 \phi_3 + \phi_4 \phi_5+ \dots +
  \phi_{n-2}\phi_{n-1})$\ ;
\item[(ii)] $E_6: I_3 = \det( \phi )$, with $ \phi$ a $3\times 3$
  matrix;
\item[(iii)] $E_7: I_3 = {\mbox{Pfaffian}}( \phi )$, with 
$ \phi$ an antisymmetric $6\times 6$ matrix;
\item[(iv)] $E_8: I_3 = \phi^3\vert_{\bf 1}$, with 
$\phi$ a {\bf 27} representation of $E_6$ and $I_3$ the singlet in 
the cubic power of  {\bf 27};
\item[(v)] $F_4: I_3 =  \det( \phi )/2$, with $ \phi$ a symmetric $3\times 3$
  matrix;
\item[(vi)] $G_2: I_3 =  \phi^3/(3\sqrt{3})$, with $\phi$ a 
single real number.
\end{enumerate}
Furthermore, for each of these cases, one can find a measure factor $\mu$
such that the Fourier transform of $\mu \exp(i I_3(\phi)/\phi_0)$
in all $\phi_0, \phi_i$ variables is exactly given by its 
one-loop saddle point 
approximation, and again of the type above (see \cite{ekp} for details).

The assignment of a simple Lie group to each cubic form in the list
above may seem arbitrary, yet it stems from standard relations
between Jordan algebras, Jordan triple systems and 
the Tits-Kantor-Koecher construction of Lie algebras
(physicists may read e.g. \cite{Gunaydin:2000xr}). 
Recall that symplectic groups $Sp(2n)$
admit a so-called metaplectic or oscillator representation in the
space of functions $\varphi(x_i)$ in 
$n$ variables, with infinitesimal generators 
\begin{equation}
\label{efh}
E_{ij}=\partial_i \partial_j\ ,\quad F_{ij}=x_i x_j\ , \quad
H_{ij}=\partial_i x_j + x_j \partial_i 
\end{equation}
where $x_i, \partial_j$ satisfy the Heisenberg relations 
$[\partial_i, x_j]=\delta_{ij}$. The $F$ generators form
a commuting subalgebra which can be exponentiated into
multiplication by a Gaussian character, $\chi_f=\exp( i f_{ij} x^i x^j)$.
On the other hand, the Weyl reflection $W$ that exchanges
$E_{ij}$ and $F_{ij}$ is the Fourier transform with respect
to all $x_i$. The closure of (\ref{efh}) can ultimately be
related to the fact that the Gaussian character $\chi_f$ is covariant
under Fourier transform $W$. This construction yields the unitary
representation of $Sp(2n)$ (or its double cover, rather) of minimal 
dimension, also called the minimal representation of $Sp(2n)$.
A similar construction can be carried out for any simple Lie group $G$
in the split real form (see \cite{ks}, \cite{kpw} for details). 
Physically, it amounts to quantize a
quantum mechanical system whose phase space is the coadjoint
orbit of a nilpotent (in particular non diagonalizable) element of
$G$. The generators acting on the wave functions take a form similar
to (\ref{efh}) above, and their closure is ultimately  related 
to the invariance of the cubic character
$\exp( i I_3(\phi_i)/\phi_0)$. This
construction has been recently used to produce new quantum mechanical
models with enhanced conformal symmetry \cite{pw}. It is also the
first step in constructing automorphic theta series for non-symplectic
groups, which have been proposed to arise in the quantization
of the eleven-dimensional BPS supermembrane \cite{Pioline:2001jn,kpw}.

Returning to the list of admissible cubic forms (i)-(vi) above, we see
that we now have a set of non-Gaussian free ``field theories'', albeit
in finite dimension. The Wick theorem can  be
straightforwardly generalized by differentiating Eq. (\ref{eleg}) with
respect to $J_i$ on either side, and now involves ``three-pronged''
propagators, or more aptly {\it junctions},
\begin{equation}
\langle~\phi_i \phi_j \phi_k ~\rangle = \frac{i}{\hbar J_0} C_{ijk}
\end{equation} 
except for propagators involving the distinguished variable $\phi_0$.
Notice that it is no longer possible to set the source $J_0=0$ after 
differentiating, due to the poles appearing in the junction. It may however
be sensible to set it equal to the quantum $\hbar$. Under this
assumption, a ``three-pronged'' version of Wick's theorem yields e.g.
\begin{eqnarray}
\langle~\phi_1^2 \phi_2^2 \phi_3 \phi_4 ~\rangle &= & \nonumber
-\left( 4 C_{1 2 3}C_{1 2 4}
+ 2 C_{1 2 2}C_{1 3 4} +
 C_{1 1 4}C_{2 2 3} + \right. \\ && \left. \qquad \qquad 
+ C_{1 1 3}C_{2 2 4} +
 2C_{1 1 2}C_{2 3 4} \right)
\end{eqnarray}
One may therefore perturb away from this free
non-Gaussian point and carry out a standard perturbative
analysis. 

Despite these attractive features,
it should be noted that these free theories unfortunately seem to
exist  only in
finite dimension (arbitrarily large in the $D_n$ case).
Indeed, in contrast to the Gaussian case, there is no notion
of tensor product of cubic forms that would allow to construct
field theoretical models from finite dimensional building
blocks. This is perhaps related to similar issues arising in attempts
to quantize Nambu dynamics \cite{Takhtajan}, 
in turn related to the quantization of
the topological membrane \cite{top}. Indeed, Nambu dynamics replaces the
near-Gaussian action $S=\int H dt - pdq$ on the wordline of a particle
by a cubic action $S=\int H dK dt - \phi_1 d\phi_2 \wedge d\phi_3$ on the
worldsheet of a string, where $H$ and $K$ are a pair of Hamiltonians
generating the motion. It would be very interesting if a suitable
modification of the cubic topological action $\int \phi_1 d\phi_2 \wedge
d\phi_3$, possibly by introducing a new coordinate $\phi_0$, led to a free
theory along the lines of this note. This may prove an important
step in the quantization of the membrane, and of the associated
non-Abelian fivebrane.

\begin{acknowledgments}
I am grateful to D. Kazhdan for introducing me to non-Gaussian
Poisson resummation, to A. Waldron for continued collaboration
on this subject, and to K. Scharnhorst for comments 
on an earlier version of this note. It is a pleasure to 
thank the other organizers of the 
Carg\`ese 2002 ASI for an inspiring and refreshing session.

\end{acknowledgments}

\end{document}